\documentclass{iau}
\usepackage{graphicx}

\title[Simulation of CMB interferometric data] %% give here short title %%
{Simulation of the analysis of interferometric microwave background polarization data}

\author[E.F. Bunn \etal]   %% give here short author list %%
{Emory F. Bunn$^1$, Ata Karakci$^2$, Paul M. Sutter$^{3,4}$, Le Zhang$^5$,
Gregory S. Tucker$^2$, Peter T. Timbie$^5$ \and Benjamin D. Wandelt$^4$}

\affiliation{$^1$University of Richmond, USA \\ email:{\tt ebunn@richmond.edu}
\\[\affilskip]
$^2$Brown University, USA\\[\affilskip]
$^3$Ohio State University, USA\\[\affilskip]
$^4$Institut d'Astrophysique de Paris, France\\[\affilskip]
$^5$University of Wisconsin -- Madison, USA
}

\pubyear{2014}
\volume{306}  %% insert here IAU Symposium No.
\pagerange{xxx--xxx}
\jname{Statistical Challenges in 21st Century Cosmology}
\editors{A.F. Heavens, J.-L. Starck, A. Krone-Martins, eds.}
\begin{document}

\maketitle

\begin{abstract}
We present results from an end-to-end simulation pipeline
interferometric observations of cosmic microwave background
polarization. We use both maximum-likelihood and Gibbs sampling
techniques to estimate the power spectrum. In addition, we use Gibbs
sampling for image reconstruction from interferometric
visibilities. The results indicate the level to which various
systematic errors (e.g., pointing errors, gain errors, beam shape
errors, cross- polarization) must be controlled in order to
successfully detect and characterize primordial B modes as well as
other scientific goals. In addition, we show that Gibbs sampling is an
effective method of image reconstruction for interferometric data in
other astrophysical contexts.
\end{abstract}

%              % remove some space above using this command.
%\section{Introduction}

Measurement of cosmic microwave background (CMB) anisotropy has become
one of the most powerful tools in cosmology. In recent years, researchers
have built on the success of these anisotopy measurements by studying
linear polarization in the CMB. In particular, considerable
attention has been focused on the search for
``B-mode'' polarization, which has the potential to measure
the energy scale of inflation, along with probing cosmology
in a variety of other ways \cite{HuDod}.
BICEP2 has measured a B-mode polarization signal in the microwave sky
\cite{bicep}, which if confirmed will represent a major advance
in cosmology. The field eagerly awaits other measurements at different
frequencies and with different instruments.

Because the $B$-mode signal is expected to be very faint,
control of systematic errors is of paramount importance. An argument
can be made (\cite[Timbie \etal\ 2006]{Timbie06}, \cite[Bunn 2007]{Bunn07})
that interferometers provide better control of systematics than imaging
telescopes. This is one of the reasons that, for instance, the QUBIC
collaboration \cite{qubic} is constructing an instrument based on
the novel approach of bolometric interferometry.

Whether or not interferometers have \textit{better} systematic
error properties than imagers, it is clear that they have
\textit{different} sensitivity to systematics.
Interferometric systematic issues have not received as much attention
as systematics in imaging systems.
Given the importance of a robust characterization of CMB $B$ modes,
it seems worthwhile to study these effects in detail.
For these reasons, we have performed detailed simulations of
interferometric observations of CMB polarization, in order
to characterize the effects of various systematic errors
on the reconstruction of the polarization power spectra
(\cite[Karakci \etal\ 2013a]{K13a}, \cite[Karakci \etal\ 2013b]{K13b},
\cite[Zhang \etal\ 2013]{Z13}).

We have developed tools that generate visibilities for 
interferometers with arbitrary antenna placement, beam shape, etc., from
HEALPix sky maps. We can include the effects of a wide variety
of systematic errors, including beam shape and pointing errors, 
cross-polarization, gain errors, \textit{etc.} We then estimate 
power spectra from these visibilities in two independent ways, via 
maximum-likelihood estimation \cite{Hobson}and Gibbs sampling
(\cite[Larson \etal\ 2007]{Larson}, \cite[Sutter \etal\ 2012]{S12}).

\begin{figure}[t]
\centerline{\includegraphics[width=1.9in]{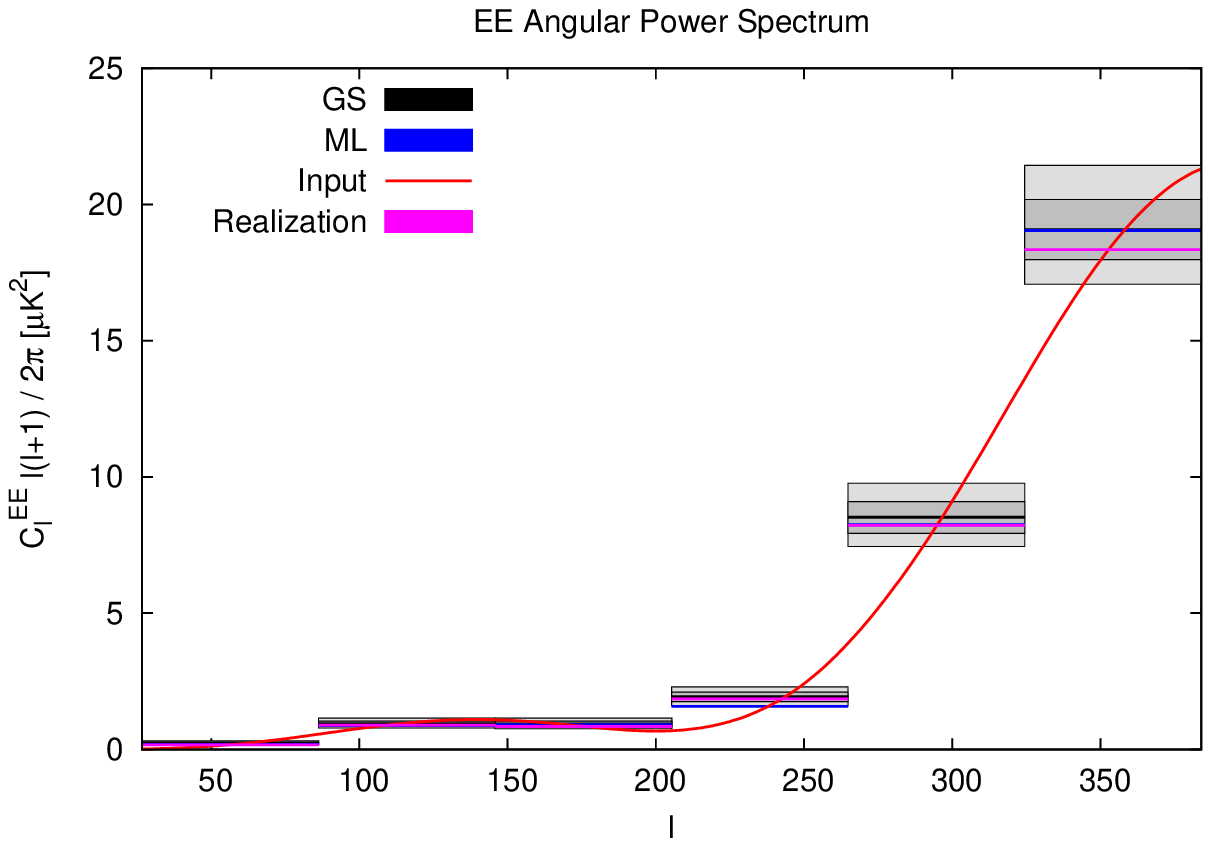}
\includegraphics[width=1.9in]{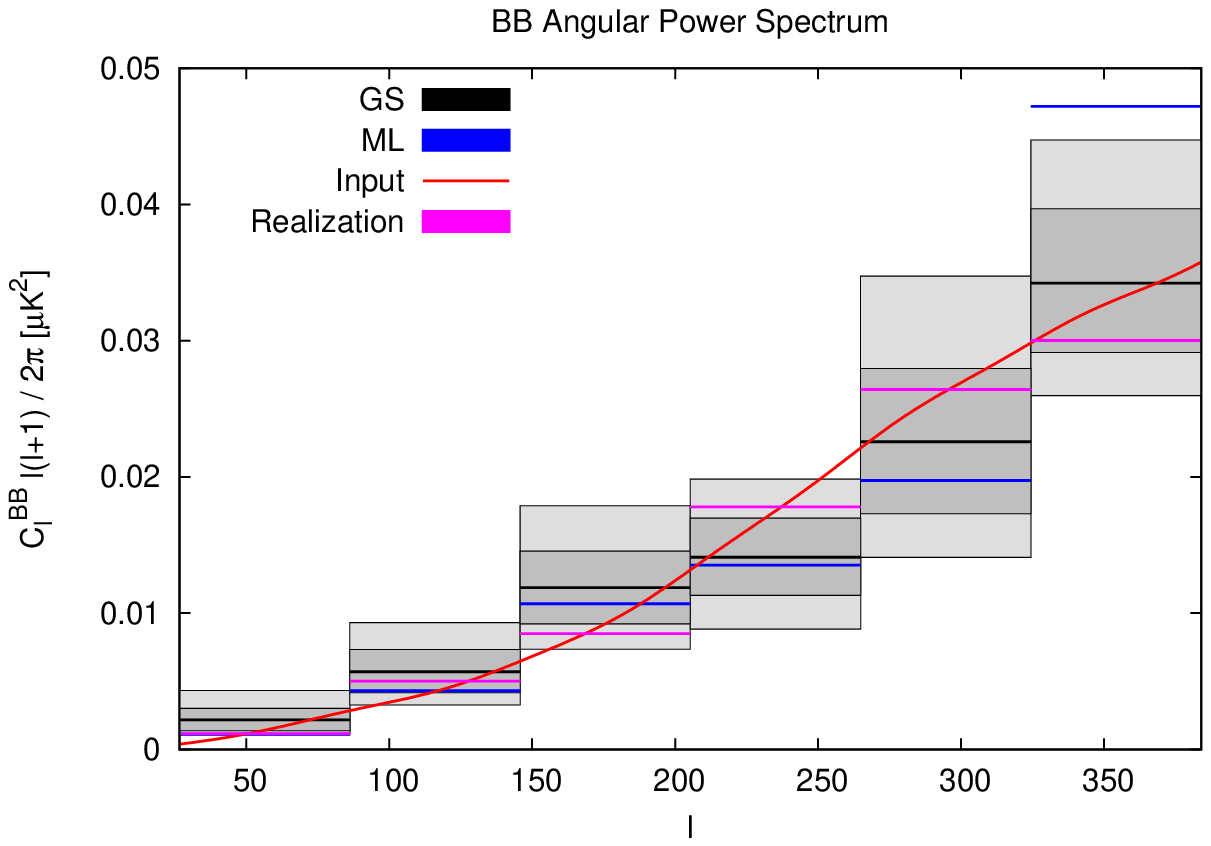}}

\centerline{
\includegraphics[width=2in]{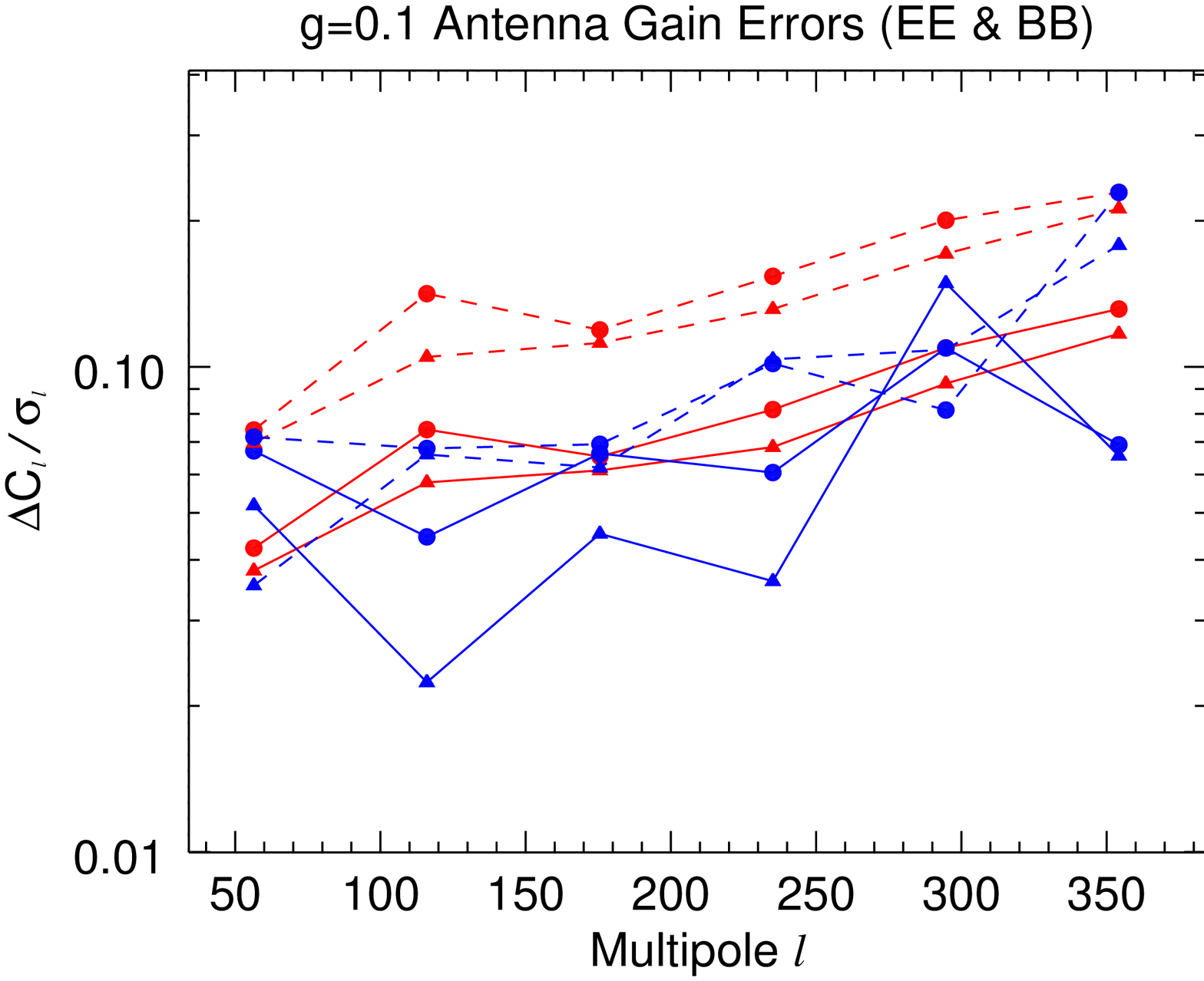}
\includegraphics[width=2in]{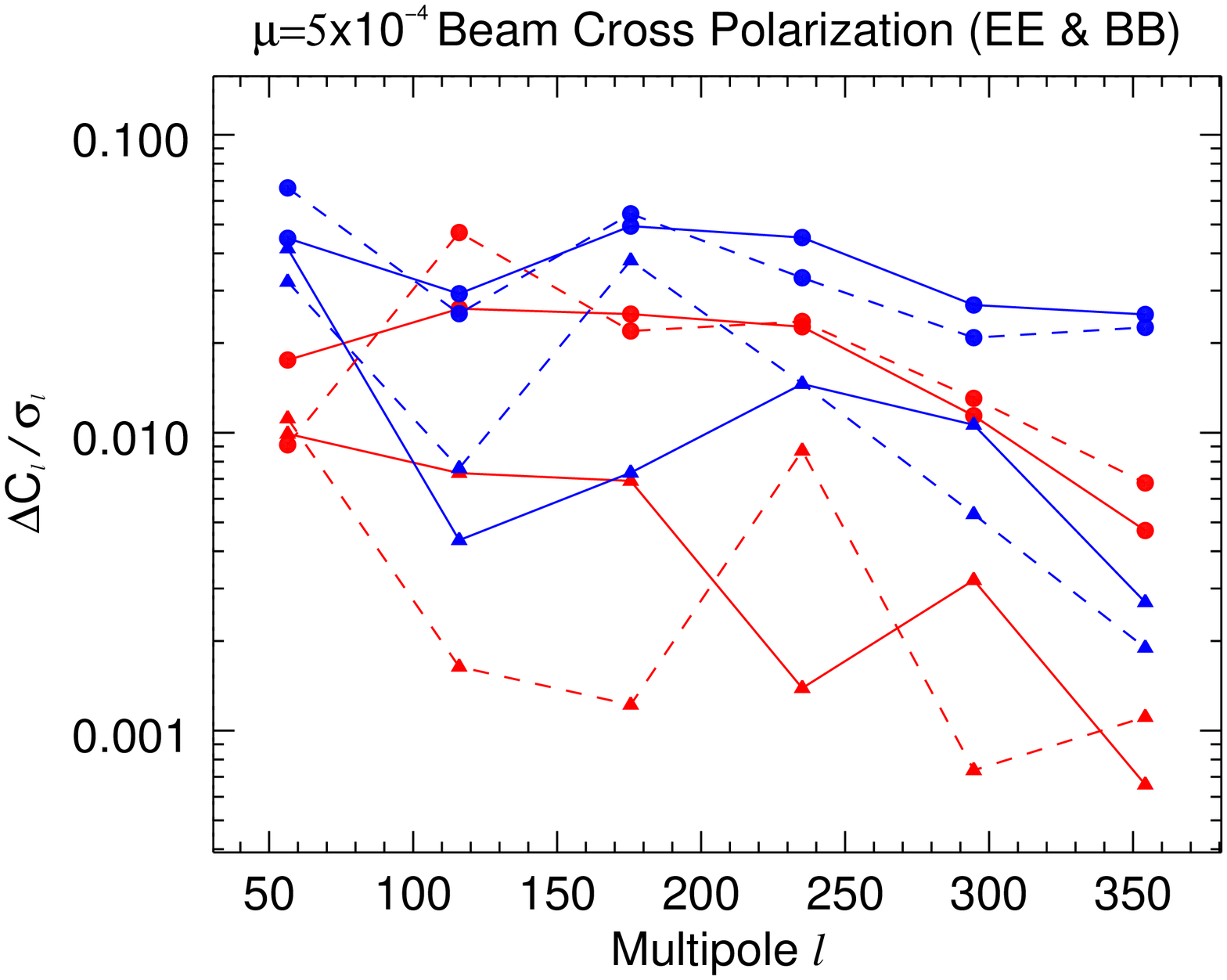}
}

\caption{Top: Mean posterior band powers obtained by Gibbs sampling (GS)
are shown in black. The maximum-likelihood (ML) 
band powers are shown in blue.
Dark and light grey indicate $1\sigma$ and $2\sigma$ uncertainties
on the Gibbs sampling results. Binned power spectra of the
signal realization and input power spectra are shown
in pink and red. \textit{Bottom}: Examples of the effects of
systematic errors, obtained by both ML (triangles) and GS (dots) are 
shown. Solid and dashed lines correspond to experiments that
interfere linear and circular polarization states respectively. 
Results are shown for the EE (red) and BB (blue) power spectra.
For further details, see \cite[Karakci \etal\ (2013b)]{K13b}.}
\label{fig:sims}
\end{figure}

Figure \ref{fig:sims} show the results of simulations that are described
in detail in \cite[Karakci \etal\ (2013a)]{K13a}, 
\cite[Zhang \etal\ (2013)]{Z13}, and \cite[Karakci \etal\ (2013b)]{K13b}.
These results are for a simulated interferometer
consisting of a 20$\times$20 close-packed array of
feedhorns with a Gaussian beam width of $5^\circ$ and separation
$D=7.89\lambda$. We include sky rotation for observations
from the South Pole, with average noise per visibility of $0.015\,\mu$K.
Figure \ref{fig:sims} shows that we correctly reconstruct the power spectra
and illustrates the effects of introducing some systematic 
errors. Order-of-magnitude agreement is found between the results
of these simulations and a simple semi-analytic approach \cite{Bunn07}.
Far more detailed results may be found in the papers cited
above.

\begin{figure}[t]
% \vspace*{-2.0 cm}
\begin{center}
 \includegraphics[width=1.3in]{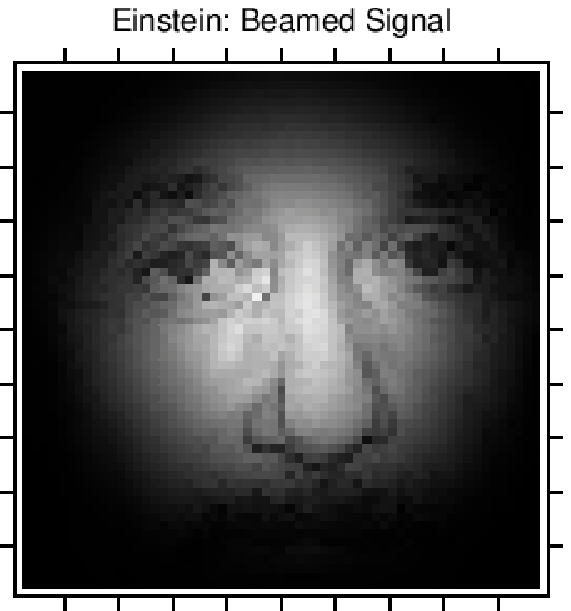} 
 \includegraphics[width=1.3in]{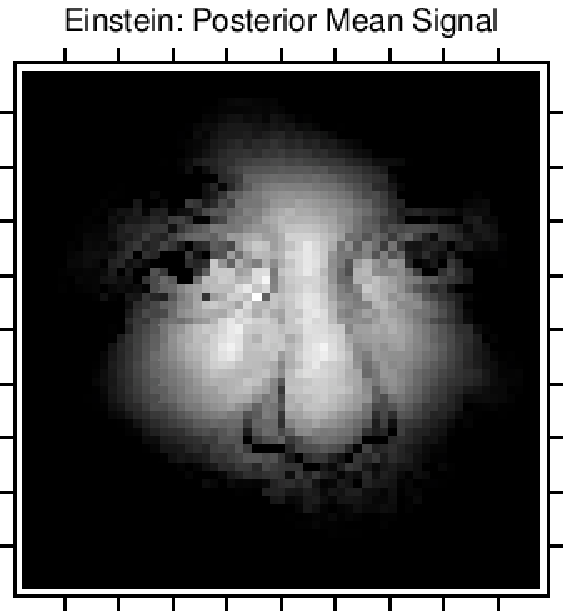} 
 \includegraphics[width=1.3in]{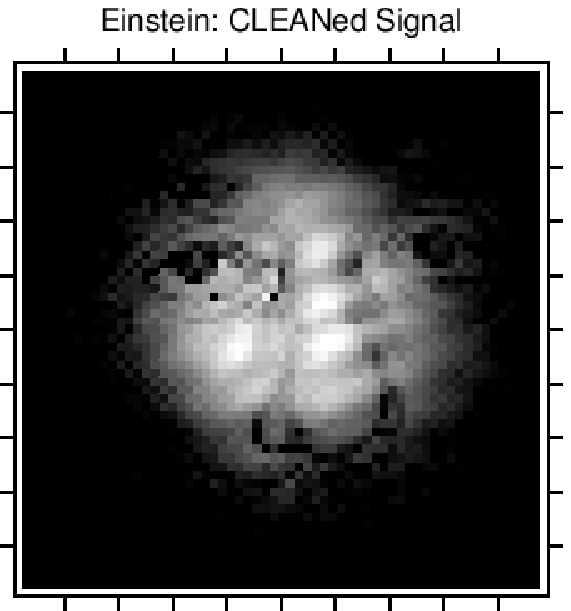} 

 \includegraphics[width=1.3in]{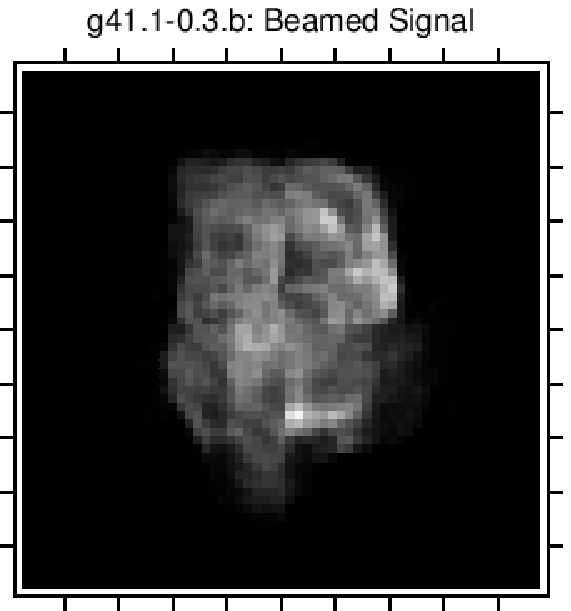} 
 \includegraphics[width=1.3in]{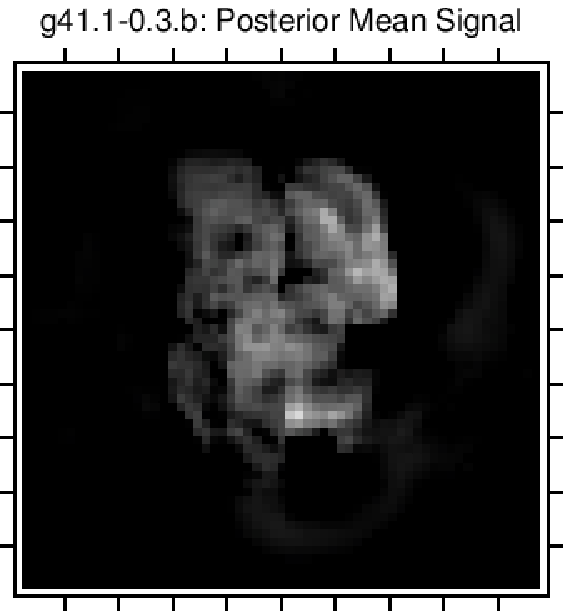} 
 \includegraphics[width=1.3in]{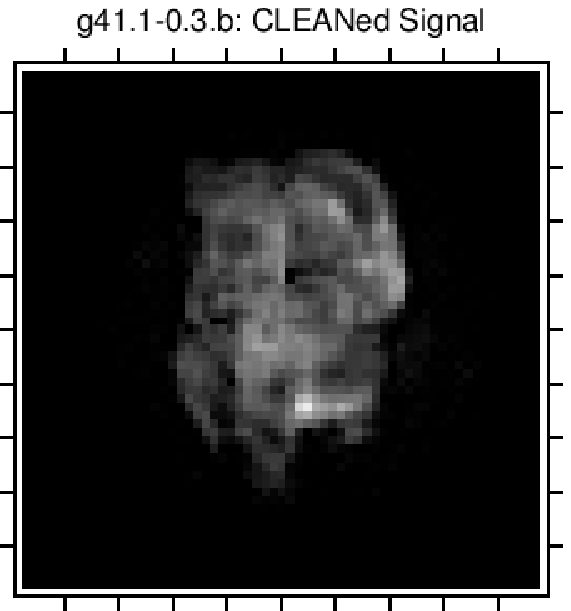}

% \vspace*{-1.0 cm}
 \caption{
Examples of image reconstruction. The left panel shows test images
from the CASA user guide, multiplied by the
primary beam of our simulated observations. 
The center panel shows the result of image
reconstruction using Gibbs sampling, and the right shows
$\ell_1$ reconstruction. Adapted from \cite[Sutter \etal\ (2014)]{S14}.
}   
\label{fig:image}
\end{center}
\end{figure}

Gibbs sampling provides simultaneous samples of the power spectrum
and the signal map. We have shown that these signal map samples provide
excellent image reconstruction from visibility data in other (non-CMB)
contexts \cite{S14}. 
Figure \ref{fig:image} shows some of the results of this work.

On the left of Figure \ref{fig:image} 
we show sample images taken from the CASA user guide
\cite{Jaeger}. These images were then ``observed'' with a simulated 
interferometer consisting of 12 randomly-placed antennas, each with a
beam size of 0.075 times the image width. We assumed 6 hours of observation
with a signal-to-noise ratio of 10 per visibility. 
The center panel shows the mean reconstructed image calculated via
Gibbs sampling, multiplied by the primary beam. The right panel shows
the result of $\ell_1$ reconstruction in the pixel basis, which has
been shown \cite{Wiaux09} to have similar performance to the
widely-used CLEAN algorithm \cite{Hogbom74}.
By a variety of quantitative measures, Gibbs reconstruction performs
better than this proxy for CLEAN reconstruction \cite{S14}.

This research was funded by the following awards from the National
Science Foundation: AST-0908902, AST-0908844, AST-0908900, AST-0908855, 
AST-0927748.

\end{document}